\documentclass{tlp}

\usepackage{harvard}
\usepackage{aopmath}

\def\i#1{\hbox{\it #1\/}}

\newcommand{\rh}{\i{head}}
\newcommand{\rp}{\i{pos}}
\newcommand{\rn}{\i{neg}}
\newcommand{\ra}{\i{atoms}}

\newcommand{\cn}[1]{{\i{Cn}({#1})}}

\newcommand{\nf}{{\i{not}\;}}
\newcommand{\rif}{\leftarrow}

\newcommand{\bu}[1]{b_{U}({#1})}

\newcommand{\tulong}[1]{{#1} \setminus \bu{#1}}

\newcommand{\eu}[2]{e_{U}\left({#1},{#2}\right)}

\newcommand{\museq}{\langle U_\alpha\rangle_{\alpha < \mu}}
\newcommand{\supr}[1]{\bigcup_{\gamma < {#1}} U_\gamma}

\title[Order-Consistent Programs are Cautiously Monotonic]
{{Order-Consistent Programs are \\
        Cautiously Monotonic}}

\author[Hudson Turner]
{ HUDSON TURNER \\
Computer Science Department\\
University of Minnesota, Duluth \\
{\tt hudson@d.umn.edu}}

\begin{document}
\bibliographystyle{dcu}

\maketitle

\begin{abstract}
Some normal logic programs under the answer set (or stable model)
semantics lack
the appealing property of
``cautious monotonicity.''  That is, 
augmenting a program with one of its consequences
may cause it to lose another of its consequences.
The syntactic condition of ``order-consistency'' was shown by Fages
to guarantee existence of an answer set.  This note establishes
that order-consistent programs are not only consistent, 
but cautiously monotonic.  From this it follows that they are
also ``cumulative.''
That is, augmenting an order-consistent
program with some of its
consequences does not alter its consequences.
In fact, as we show, its answer sets remain unchanged.
\end{abstract}

\section{Introduction} \label{sec:introduction}

The answer set (or stable model) semantics of normal logic programs
(Gelfond and Lifschitz 1988, 1991) \nocite{gel88,gel91b}
does not satisfy cautious monotonicity.  That is,
even if atoms~$a$ and~$c$ are among the consequences of a program~$P$, 
$a$~may fail to be a consequence of the program ${P \cup \{c \rif\}}$.
Here is an example due to Dix \nocite{dix91} (1991), who has published many
studies of such properties for various logic programming semantics.
\begin{eqnarray*}
 && a \rif \nf b \\
 && b \rif c, \nf a \\
 && c \rif a
\end{eqnarray*}
This program has only one answer set $\{a,c\}$, and so has $a$ and $c$
among its consequences.
When augmented
with the rule~${c \rif}$
the program gains a second answer set~$\{b,c\}$, and loses consequence~$a$.

A syntactic condition known as ``order-consistency'' \cite{sat90} was
shown by Fages (1994) \nocite{fag94}
to guarantee consistency of normal programs
under the answer set semantics.  In this note
we establish that order-consistency guarantees
another nice property: cautious monotonicity.

\begin{theorem} [Cautious Monotonicity Theorem]
If $P$ is an order-consistent program and atom~$a$ belongs to every answer
set for~$P$, then every answer set for program~${P \cup \{a\rif\}}$
is an answer set for~$P$.
\end{theorem}

All normal programs under the answer set semantics
have a property complementary to cautious monotonicity,
commonly called ``cut'': augmenting a program
with one of its consequences
cannot cause it to gain a consequence.
This is immediate, given the following easily proved fact.

\begin{fact}
If an atom~$a$ belongs to an answer set $X$ for a program $P$, then $X$ is an answer
set for program~$P \cup \{a\rif\}$.
\end{fact}

Cut and cautious monotonicity together imply another nice property,
called ``cumulativity'': augmenting a program with one of its
consequences does not alter its consequences.
Corresponding to this, 
we have the following result for
order-consistent programs.

\begin{corollary} [Cumulativity Corollary]
If an atom $a$ belongs to every answer set for an order-consistent program~$P$,
then programs $P$ and ${P \cup \{a\rif\}}$ have the same
answer sets.
\end{corollary}

Semantic properties such as
cumulativity, cut and cautious monotonicity were
originally formulated more generally for analysis of
consequence relations lacking the
classic monotonicity property \cite{gab85,mak89,kra90}.
Makinson's (1993) \nocite{mak93} handbook article 
includes a survey of such properties for nonmonotonic logics
used in AI, among them logic programming under 
the stable model (answer set) semantics.

The remainder of this note is devoted
to a proof of the Cautious Monotonicity Theorem
(and also, of course, to recalling the definitions involved in its statement).
Here is a preliminary sketch.
We first observe that adding a consequence to a 
``signed'' program does not alter its answer sets.
(This follows from results due to
Dung (1992) \nocite{dun92} and Schlipf.)
We then recall a result from \cite{lif94e}
that characterizes the
answer sets~$X$ for an order-consistent program~$P$
in terms of families of signed programs
whose answer sets correspond to a partition of~$X$.
In the proof we establish in addition that
if an atom $a$ is a consequence of order-consistent program~$P$, then
$a$ is a consequence of the corresponding member of each of the families of
signed programs.  It follows that adding rule~${a\rif}$ to~$P$, and so
to the corresponding member of each of the families of signed programs,
does not affect the answer sets for the members of the families of
signed programs.  We can then conclude, by the Splitting Sequence Theorem
of \cite{lif94e}, that each answer set for~${P \cup \{a\rif\}}$ is an answer
set for~$P$.

\section{Normal Logic Programs} \label{sec:programs}

Begin with a set of symbols called \emph{atoms}.
A \emph{rule} consists of three parts: an atom called the \emph{head},
and two finite sets of atoms---the set of \emph{positive subgoals}
and the set of \emph{negated subgoals}.
The rule with head~$a$,
positive subgoals~${b_{1},\ldots,b_{m}}$ and
negated subgoals~${c_{1},\ldots,c_{n}}$ is typically written
\begin{eqnarray*}
 a \rif b_{1},\dots,b_{m}, 
 \nf c_{1},\dots,\nf c_{n}.
\end{eqnarray*}
We denote the three parts of a rule~$r$ by $\rh(r)$, $\rp(r)$ and
$\rn(r)$; $\ra(r)$ stands for ${\{\rh(r)\} \cup \rp(r) \cup \rn(r)}$.

A \emph{program} is a set of rules.   For any program~$P$, by $\ra(P)$ we
denote the union of the sets $\ra(r)$ for all ${r \in P}$; the atoms in
this set are said to \emph{occur} in~$P$.

A program $P$ is \emph{positive} if, for every rule ${r \in P}$,
${\rn(r) = \emptyset}$.  The notion of an answer set is first defined
for positive programs, as follows.
A set $X$ of atoms is \emph{closed} under a positive program~$P$
if, for every rule ${r \in P}$ such that ${\rp(r) \subseteq X}$,
${\rh(r) \in X}$.
The \emph{answer set}
for a positive program~$P$ is the least set of atoms closed
under~$P$.

Now let $P$ be an arbitrary program and $X$ a set of atoms.  For each
rule ${r \in P}$ such that ${\rn(r) \cap X = \emptyset}$, let $r'$ be the rule
defined by
\begin{eqnarray*}
\rh(r')=\rh(r)\,,\; \rp(r')=\rp(r)\,,\; \rn(r')=\emptyset\,.
\end{eqnarray*}
The positive program consisting of all rules~$r'$ obtained in this way is the
\emph{reduct} of~$P$ relative to~$X$, denoted by~$P^X$.
We say $X$ is an \emph{answer set} for~$P$ 
if $X$ is the answer set for~$P^X$.

A program is \emph{consistent} if it has an answer set.
An atom is a \emph{consequence} of a program~$P$ if it belongs to all
answer sets for~$P$.  We write $\cn{P}$ to denote the set of all consequences
of~$P$.

We'll want an auxiliary notion, related to the well-founded semantics
of logic programs \cite{van91}.  
For any program~$P$, let $\Gamma_P$ be the operator that
maps a set~$X$ of atoms to the answer set for~$P^X$.  Clearly, the
answer sets for~$P$ are exactly the fixpoints of $\Gamma_P$. It is
well-known that $\Gamma_P^2$ is a monotone operator whose least fixpoint,
which we'll denote by~$\i{WF}(P)$,
is exactly the set of atoms true in the well-founded model of~$P$.

\section{Signed Programs are Cautiously Monotonic} \label{sec:cautious}

A program~$P$ is \emph{cautiously monotonic} if, for
all ${a \in \cn{P}}$,
\begin{eqnarray*}
\cn{P} \subseteq \cn{P \cup \{ a \rif \}}\,.
\end{eqnarray*}

A program~$P$ is \emph{cumulative} if, for
all ${a \in \cn{P}}$,
\begin{eqnarray*}
\cn{P} = \cn{P \cup \{ a \rif \}}\,.
\end{eqnarray*}

We are interested in a stronger property: 
for all ${a \in \cn{P}}$, programs~$P$ and ${P \cup \{ a \rif \}}$
have the same answer sets.

To see that this is indeed a stronger property, notice that adding
the rule~${a \rif}$ to program $\{ a \rif \nf a \}$ changes its answer sets, but
not its consequences.

The notion of a ``signing'' of a program is due to Kunen (1989). \nocite{kun89}
A program~$P$ is \emph{signed}
if there is a set~$S$ of atoms such that,
for every rule ${r \in P}$,
\begin{itemize}
 \item if $\rh(r) \in S$ then $\rp(r)\subseteq S$ and $\rn(r)\cap S = \emptyset$\,,
 \item if $\rh(r) \notin S$ then $\rp(r)\cap S = \emptyset$ and $\rn(r)\subseteq S$\,.
\end{itemize}

The following program~$P_1$ is signed.
\begin{eqnarray*}
 && a \rif \nf b \\
 && b \rif \nf a
\end{eqnarray*}
Take $S= \{a\}$, for instance.

\begin{lemma} [Signing Lemma]
For any signed program~$P$ and ${a \in \cn{P}}$,
programs~$P$ and ${P \cup \{a \rif\}}$
have the same answer sets.
\end{lemma}

This is immediate, given the following two results.

\begin{proposition}
\cite{dun92}
For any signed program~$P$, ${\cn{P} = \i{WF}(P)}$.
\end{proposition}

\begin{proposition}
(Schlipf, personal communication)
For any program~$P$ and ${a \in \i{WF}(P)}$,
programs~$P$ and ${P \cup \{ a \rif \}}$
have the same answer sets.
\end{proposition}

Proposition~1 follows from a stronger result in \cite{dun92}.
Proposition~2 is apparently widely known, and
plays a significant role in automated systems
for answer set programming.

\section{Order-Consistent Programs}    \label{sec:order}

For any program~$P$ and atom~$a$, $P_a^{+}$ and $P_a^{-}$ are the
smallest sets of atoms such that $a\in P^{+}_a$ and,
for every rule $r\in P$,
\begin{itemize}
\item
if $\rh(r)\in P^{+}_a$ then $\rp(r)\subseteq P^{+}_a$ and $\rn(r)\subseteq P^{-}_a$,
\item
if $\rh(r)\in P^{-}_a$ then $\rp(r)\subseteq P^{-}_a$ and $\rn(r)\subseteq P^{+}_a$.
\end{itemize}
Intuitively, $P^{+}_a$ is the set of atoms on which atom~$a$ depends positively in~$P$,
and $P^{-}_a$ is the set of atoms on which atom~$a$ depends negatively in~$P$.

A \emph{level mapping} is a function from atoms to ordinals.

A program~$P$ is \emph{order-consistent} if there
is a level mapping~$\lambda$ such that ${\lambda(b) < \lambda(a)}$
whenever ${b\in P^{+}_a \cap P^{-}_a}$.  That is, if $a$ depends both positively and
negatively on~$b$, then $b$ is mapped to a lower stratum.

\begin{theorem} [Fages' Theorem]
\cite{fag94}
Order-consistent programs are consistent.
\end{theorem}

The following program~$P_2$ is order-consistent.
\begin{eqnarray*}
 && a \rif \nf b \\
 && b \rif \nf a \\
 && c \rif a \\
 && c \rif b
\end{eqnarray*}
Consider, for example, the level mapping ${\lambda(a) = \lambda(b) = 0}$,
${\lambda(c) = 1}$.

Clearly every signed program is order-consistent.  As program~$P_2$ illustrates,
the converse does not hold.

\section{Call-Consistent Programs are not Cautiously Monotonic}

For finite programs, order-consistency is equivalent to a well-known,
simpler condition:
a program~$P$ is \emph{call-consistent} if for
all ${a \in \ra(P)}$, ${a \notin P^{-}_a}$.  That is, no atom depends negatively
on itself.

The following (infinite) program is call-consistent, but not order-consistent.
\begin{eqnarray*}
 && a_m \rif \nf c, \nf a_n \qquad (0 \leq m < n)
\end{eqnarray*}
This program has no answer set, so $c$ and $a_0$ are among its consequences.
Adding the rule~${c \rif}$
produces a single answer set ${\{ c \}}$ and
thus eliminates consequence~$a_0$.
This shows that not all call-consistent programs are cautiously monotonic.

One may wonder at this point if all \emph{consistent} call-consistent programs
are cautiously monotonic.  Consider adding the following rules to
the previous example.
\begin{eqnarray*}
 && c \rif a \\
 &&  a \rif \nf b \\
 &&  b \rif \nf a
\end{eqnarray*}
The resulting program has a single answer set ${\{ a,c \}}$.
Adding the
rule~${c \rif}$ yields a second answer set ${\{ b,c \}}$.

\section{Splitting Sequences}  \label{sec:splitting}

In order to ``decompose'' an order-consistent program into a family of
signed programs and reason about the result, we need some machinery.
The definitions given in this section and the next
simplify (slightly) those from \cite{lif94e},
which applied also to non-normal programs (with classical negation and disjunction).

A \emph{splitting set} for a program~$P$ is any set~$U$ of atoms such that,
for every rule~${r \in P}$, if ${\rh(r) \in U}$ then
${\ra(r) \subseteq U}$.  

It is clear that for any program~$P$, both $\emptyset$ and $\ra(P)$ are splitting
sets.  For program~$P_2$ from Section~\ref{sec:order}, another splitting set
is~${\{a,b\}}$.

Let $U$ and $X$ be sets of atoms and $P$ a program.
The set of rules~${r \in P}$
such that ${\ra(r) \subseteq U}$ is denoted by $\bu{P}$.
For each rule ${r\in \tulong{P}}$ such that ${\rp(r) \cap U \subseteq X}$ and
$\rn(r) \cap X = \emptyset$, take the rule $r'$ defined by
\begin{eqnarray*}
\rh(r')=\rh(r)\,,\; \rp(r')=\rp(r) \setminus U\,,\; \rn(r')=\rn(r) \setminus U\,.
\end{eqnarray*}
The program consisting of all rules $r'$ obtained in this way is denoted
by~$\eu{P}{X}$.

For example, if ${U = \{a,b\}}$ then $b_U(P_2)$ is exactly the signed program~$P_1$
considered previously, and ${e_U(P_2,\{a\}) = \{ c \rif \} = e_U(P_2,\{b\})}$.

A \emph{$($transfinite$)$ sequence} is a family whose index set is an initial
segment of ordinals, $\{\alpha : \alpha < \mu\}$.
A sequence $\museq$ of sets is \emph{monotone}
if ${U_\alpha \subseteq U_\beta}$ whenever ${\alpha < \beta}$,
and \emph{continuous} if,  for each limit ordinal~${\alpha < \mu}$,
${U_\alpha = \supr{\alpha}}$.

A \emph{splitting sequence} for a program~$P$ is a nonempty, monotone, continuous
sequence $\museq$ of splitting sets for~$P$ such that
$\bigcup_{\alpha < \mu} U_\alpha=\ra(P)$.

For example, ${\langle \{a,b\},\{a,b,c\}\rangle}$
is a splitting sequence for program~$P_2$.

Let ${U = \museq}$ be a splitting sequence
for a program~$P$.  A \emph{solution} to~$P$ (with respect to~$U$) is a
sequence ${\langle X_\alpha\rangle_{\alpha < \mu}}$ of sets of atoms
such that
\begin{itemize}
\item $X_0$ is an answer set for~$b_{U_0}(P)$\,,
\item for any $\alpha$ such that ${\alpha+1 < \mu}$,
      $X_{\alpha+1}$ is an answer set for
     \begin{eqnarray*}
      e_{U_\alpha}\left(b_{U_{\alpha+1}}(P),
      \bigcup_{\gamma\leq\alpha}X_\gamma\right),
     \end{eqnarray*}
\item for any limit ordinal $\alpha < \mu$, $X_\alpha = \emptyset$\,.
\end{itemize}

Notice, for example, that program~$P_2$ has two solutions with
respect to splitting sequence ${\langle \{a,b\},\{a,b,c\}\rangle}$: 
${\langle \{a\},\{c\}\rangle}$ and ${\langle \{b\},\{c\}\rangle}$.
They correspond to the two answer sets for~$P_2$, as described
in the following general theorem.

\begin{theorem} [Splitting Sequence Theorem]
\cite{lif94e}
Let ${U=\museq}$ be a splitting sequence for a program~$P$.
A set $X$ of atoms is an answer set for~$P$ if and only if
\begin{eqnarray*}
  {X=\bigcup_{\alpha < \mu} X_\alpha}
\end{eqnarray*}
for some solution
${\langle X_\alpha\rangle_{\alpha < \mu}}$
to~$P$ with respect to~$U$.
\end{theorem}

Let ${U = \museq}$ be a splitting sequence for a program~$P$.
A sequence
${\langle X_\alpha\rangle_{\alpha < \mu}}$ of sets of atoms
``decomposes'' $P$ into the following family of programs.
\begin{eqnarray}
 && b_{U_0}(P)       \label{eq:bu} \\
 && e_{U_\alpha}\left(b_{U_{\alpha+1}}(P),
     \bigcup_{\gamma\leq\alpha}X_\gamma\right) 
     \qquad(\alpha+1<\mu) \label{eq:eu}
\end{eqnarray}
Every atom occurring in~(\ref{eq:bu}) belongs to~$U_0$,
and for every ${\alpha + 1 < \mu}$, every atom occurring in~(\ref{eq:eu})
belongs to~${U_{\alpha+1} \setminus U_\alpha}$.
Consequently, the members of any solution are
answer sets for a family of programs no two of which
have an atom in common.

\section{Signed Components of Order-Consistent Programs}  

We are interested in the syntactic form of the programs~(\ref{eq:bu})
and (\ref{eq:eu}) whose answer sets can be members of a solution
to an order-consistent program~$P$.
It is clear that each rule of each of these programs is obtained from
a rule of~$P$ by removing some of its subgoals.
A more specific claim can be made using the following terminology.

For any program~$P$ and set $X$ of atoms, let $\i{rm}(P,X)$ be the
program obtained from~$P$
by removing from each of the rules of~$P$
all subgoals, both
positive and negated, that belong to~$X$.
For any program~$P$ and splitting sequence ${U=\museq}$ for~$P$, the
programs
\begin{eqnarray*}
 && b_{U_0}(P)\,,\\
 && \i{rm}\left(b_{U_{\alpha+1}}(P) 
      \setminus b_{U_\alpha}(P),U_\alpha\right)
    \qquad(\alpha+1<\mu)
\end{eqnarray*}
will be called the \emph{$U$-components} of~$P$.

It is easy to see that, for any set~$X$ of atoms,
\begin{eqnarray*}
e_{U_\alpha}(b_{U_{\alpha+1}}(P),X) \subseteq
  \i{rm}(b_{U_{\alpha+1}}(P) \setminus b_{U_\alpha}(P),U_\alpha)\,.
\end{eqnarray*}
Consequently, each of the programs~(\ref{eq:bu})
and~(\ref{eq:eu}) is a subset of the corresponding
$U$-component of~$P$.

In \cite{lif94e} we showed that 
a program is stratified if and only if
it has a splitting sequence~$U$ such that all $U$-components are
positive.  We also established the following
characterization of order-consistent programs.

\begin{proposition}
 \cite{lif94e}
A program~$P$ is order-consistent if and only if it has a splitting sequence~$U$
such that all \hbox{$U$-components} of~$P$ are signed.
\end{proposition}

For example, if ${U = \langle \{a,b\},\{a,b,c\}\rangle}$, then the 
$U$-components of~$P_2$ are the signed programs~$P_1$ and ${\{ c\rif \}}$.

As discussed in \cite{lif94e}, Proposition~3
and the Splitting Sequence Theorem can be used to derive Fages'
Theorem from a similar---and easier---result for signed programs.
Below they are used instead in the proof that order-consistent programs
are cautiously monotonic.

\section{Proof of Cautious Monotonicity Theorem}

\begin{quote}\normalsize 
    \noindent{\it Restatement of Theorem 1.}
If $P$ is an order-consistent program and ${a \in \cn{P}}$,
then every answer set for program~${P \cup \{a\rif\}}$ is an answer set for~$P$.
\end{quote}

\begin{proof}
Assume $P$ is order-consistent and $a \in \cn{P}$.
Let $X$ be an answer set for~$P \cup \{a\rif\}$.
Since $P$ is order-consistent, so is $P \cup \{a\rif\}$.
By Proposition~3, there is a splitting sequence ${U = \museq}$
for~$P \cup \{a\rif\}$ such that
all $U$-components of $P \cup \{a\rif\}$ are signed.
Notice that $U$ is also a splitting sequence for~$P$, and
that all $U$-components of $P$ are signed as well.
By the Splitting Sequence Theorem, there is a solution
$\langle X_\alpha\rangle_{\alpha < \mu}$ to $P \cup \{a\rif\}$
with respect to $U$ such that ${X = \bigcup_{\alpha < \mu} X_\alpha}$.
We complete the proof by showing that ${\langle X_\alpha\rangle_{\alpha < \mu}}$
is a solution to $P$ with respect to $U$.  (From this it follows, 
again by the Splitting
Sequence Theorem, that $X$ is an answer set for~$P$.)

Observe that any splitting sequence can be ``extended'' by inserting $\emptyset$
at its beginning.  That is, since $\museq$ is a splitting sequence for~$P$
and $P \cup \{a\rif\}$,
so is ${\langle U'_\alpha\rangle_{\alpha < \mu+1}}$, where 
\begin{itemize}
 \item $U'_0 = \emptyset$,
 \item for all natural numbers $n$ such that $n+1 < \mu$, $U'_{n+1}=U_n$,
 \item for all ordinals $\alpha$ such that $\omega \leq \alpha < \mu$,
   $U'_{\alpha}=U_\alpha$,
 \item $U'_\mu = \ra(P)$.
\end{itemize}
Notice that since all $U$-components of $P$ and $P \cup \{a\rif\}$ are signed,
so are all $U'$-components.
For convenience then, we will assume,
without loss of generality, that $U_0 = \emptyset$.
Under this assumption, any atom that occurs in $P$ belongs to
one of the sets $U_{\alpha+1} \setminus U_\alpha$ (for some $\alpha$
such that $\alpha+1 < \mu$).

Let $\alpha$ be such that $a \in U_{\alpha+1} \setminus U_\alpha$.
For all $\beta+1<\mu$ such that $\beta \neq\alpha$,
\begin{eqnarray*}
{e_{U_\beta}\left(b_{U_{\beta+1}}(P),
\bigcup_{\gamma\leq\beta}X_\gamma\right)}
 & = &
{e_{U_\beta}\left(b_{U_{\beta+1}}(P \cup \{a\rif\}),
\bigcup_{\gamma\leq\beta}X_\gamma\right)}\,.
\end{eqnarray*}
Hence, we can show that $\langle X_\alpha\rangle_{\alpha < \mu}$ is a solution
to $P$ with respect to $U$ simply by showing that $X_{\alpha+1}$ is
an answer set for
\begin{eqnarray}
{e_{U_\alpha}\left(b_{U_{\alpha+1}}(P),
\bigcup_{\gamma\leq\alpha}X_\gamma\right)}\,.
  \label{signed-part}
\end{eqnarray}
We will do this by showing that
(\ref{signed-part})
has the same answer sets as
\begin{eqnarray*}
{e_{U_\alpha}\left(b_{U_{\alpha+1}}(P \cup \{a\rif\}),
\bigcup_{\gamma\leq\alpha}X_\gamma\right)}\,.
\end{eqnarray*}

First, notice that the latter program is the same as
\begin{eqnarray*}
e_{U_\alpha}\left(b_{U_{\alpha+1}}(P),
\bigcup_{\gamma\leq\alpha}X_\gamma\right) \cup \{a\rif\}\,.
\end{eqnarray*}
So it is enough to show that adding the rule~${a\rif}$
to~(\ref{signed-part}) does not affect its answer sets.
Since (\ref{signed-part})~is a signed program, we can use
the Signing Lemma: it remains only to show that atom~$a$ is among
the consequences of~(\ref{signed-part}).

Take ${V_0 = U_\alpha}$, ${V_1 = U_{\alpha+1}}$ and
${V_2 = \ra(P)}$,
The sequence
${V = \langle V_0,V_1,V_2 \rangle}$
is a splitting sequence for~$P$.
We construct a solution to~$P$ with respect to~$V$ as follows.
Take ${Y_0 = \bigcup_{\gamma\leq\alpha}X_\gamma}$.
It is straightforward, using the Splitting
Sequence Theorem, to verify that $Y_0$~is
an answer set for~$b_{V_0}(P)$.
Notice that ${e_{V_0}\left(b_{V_1}(P),Y_0\right)}$ is
exactly the program~(\ref{signed-part}).
Since (\ref{signed-part})~is signed, it is consistent.
Let $Y_1$ be one of its answer sets.
Since $P$~is order-consistent, so is
${e_{V_1}(b_{V_2}(P),Y_0 \cup Y_1)}$, and, by Fages' Theorem, 
it too is consistent.  Let $Y_2$ be
one of its answer sets.
By this construction, the sequence
${\langle Y_0,Y_1,Y_2 \rangle}$ is a solution to~$P$ with respect to~$V$.
By the Splitting Sequence Theorem, ${Y = Y_0 \cup Y_1 \cup Y_2}$
is an answer set for~$P$.  Since ${a \in \cn{P}}$, ${a \in Y}$.
It follows that $a \in Y_1$.  And since $Y_1$ was an arbitrarily chosen
answer set for~(\ref{signed-part}),
we conclude that $a$~is among its consequences.
\end{proof}

\section*{Acknowledgements}

Thanks to T.C.\ Son for hinting at this question, and
to Michael Gelfond for reminding me of a related conjecture
in \cite{bar94a}.  Thanks also to Son, Michael, and Vladimir Lifschitz
for comments on an earlier draft, and to J\"{u}rgen Dix and
John Schlipf for helpful correspondence.

\end{document}